\documentclass[amsmath,superscriptaddress,amssymb,aps,prd,floatfix,twocolumn]{revtex4-2}
\usepackage{amsmath}
\usepackage{amssymb}
\usepackage{amsfonts}
\usepackage{array}
\usepackage{babel}
\usepackage{graphics}
\usepackage{graphicx}
\usepackage{lipsum}
\usepackage{mathrsfs}
\usepackage{multirow}
\usepackage{physics}
\usepackage{caption}
\usepackage{textgreek}
\usepackage{subcaption}
\usepackage{notoccite}

\usepackage{ulem}
\usepackage[bookmarks=false,hidelinks]{hyperref}

\begin{document}
\title{Motion of test particles around an Einstein-dilaton-Gauss-Bonnet black hole in a uniform magnetic field}

\author{Romel M. Vargas}
\email{rvargas@uni.edu.pe}
\affiliation{Universidad Nacional de Ingeniería, Lima, Perú}
\author{M. A. Cuyubamba}
\email{mcuyubamba@usil.edu.pe}
\affiliation{Grupo de Investigación en Física, Facultad de Ingeniería, Universidad San Ignacio de Loyola, Lima, Perú}

\begin{abstract}
In this work, we study the motion of a magnetized particle orbiting a static and spherically symmetric black hole immersed in an external asymptotically uniform magnetic field in Einstein-dilaton-Gauss-Bonnet gravity. Similar to the Schwarzschild case, the magnetic interaction creates a region that allows circular stable orbits near the photonic sphere, however, we show that regions in which there are no allowed solutions for circular orbits appear before the marginal stability was reached for weak magnetic interaction. The regions of allowed stable circular orbits were calculated for different values of the dilaton-Gauss-Bonnet coupling $p$ and magnetic coupling parameter $\beta$, concluding that the increase of $p$ reduces the regions of stable circular orbits. The calculations were carried out using numerical black hole solutions and were compared with an analytical approximation with an error below $5\%$ for $p<0.25$.
\end{abstract}

\maketitle
\section{Introduction}

The first black hole solution arose shortly after Einstein presented his theory of general relativity. Since then, the study of black holes has been important to understanding the nature of our universe. They give us an insight into the behavior of gravity in extreme conditions, as well as playing a key role in the shaping of galaxies. 

Despite the success of general relativity in explaining the nature of astronomical observations, there are still some fundamental questions left unsolved up to this date, such as the black hole singularity, nature of dark matter and dark energy, the non-renormalization problems or the incompatibility between general relativity and quantum mechanics, which indicate that general relativity is not a complete theory of gravity. For this reason, alternative models to Einstein's gravity have been proposed through the past century, with hopes of solving all these problems. One of these modifications is the Einstein-dilaton-Gauss-Bonnet (EdGB) gravity, which arises from the heterotic string theory \cite{metsaev1987,GROSS198741}, as a kind of minimal effective theory and that belongs to the class of Horndeski gravity \cite{Horndeski:1974wa,Kobayashi:2019hrl}. In the framework of EdGB, it introduces a scalar dilaton field into the action non-minimally coupled with the Gauss-Bonnet term, which is quadratic order of curvature. Such a coupling allows that this modification to the Einstein-Hilbert action bypasses Bekenstein's no-scalar-hair theorem \cite{kanti1996, bekenstein1995}. In four dimensions, the Gauss-Bonnet term alone is invariant and does not contribute to the Einstein field equations, however, it is circumvented when it is non-minimally coupled with a scalar dilaton field. 

EdGB black hole has gained interest of many physicists as an alternative model in recent years \cite{Pani:2009wy}, finding multiple numerical solutions \cite{kleihaus2011rotating, yunes2011nonspinning, pani2011slowly, ayzenberg2014,Kleihaus:2015aje,Maselli:2015tta}, analytical approximated solution \cite{kokkotas2017} and recently performing inspiral-merger-ringdown waveform models \cite{Okounkova:2020rqw,Carson:2020ter,Julie:2024fwy}. Additional aspects like the ringdown signal from the quasinormal modes, stability analysis, black hole shadows, among others \cite{Blazquez-Salcedo:2016enn,Blazquez-Salcedo:2017txk,cunha2017shadows, Okounkova:2019zep,Ripley:2019irj,Konoplya:2019hml,Pierini:2021jxd,Pierini:2022eim,Blazquez-Salcedo:2024oek,Chung:2024vaf}, were extensively studied in the last years. Nevertheless, it has proven to be difficult to test this theory due to the similarity of its effects with general relativity within the observables that are on our reach. For instance, shadows tend to be just around $1\%$ smaller to that of Kerr black holes \cite{cunha2017shadows}, and it is possible to make a good fit of the spectrum of accreting black holes in EdGB with respect to Kerr black holes \cite{zhang2017testing}, making them difficult to distinguish.

The fact that not even light can escape their gravitational attraction within their event horizon has made it difficult to have direct observations of black holes. For this reason, these objects are studied through their interactions with other celestial bodies, and the matter surrounding them. One of the characteristic features of black holes is their innermost stable circular orbit (ISCO) which defines the inner border of the accretion disk, making it important to interpret observations \cite{page1974}. For spinning black holes, the ISCO radius is different for prograde or retrograde orbits, and it varies with its spin parameter \cite{lee2017, wei2018}. Regardless, the spin of a black hole is not only factor that affects the ISCO radius. It is known that a the electric charge of a black hole will push away the ISCO radius for neutral particles, independent of the sign of the black hole's charge, while a magnetic charge will pull it closer \cite{alzahrani2022}. Furthermore, Zajacek et al found that galactic center super massive black holes (SMBH) can obtain a stable charge due the difference in mass between electrons and protons in the surrounding plasma, and the rotation of the SMBH coupled with an external magnetic field \cite{zajavcek2018}. In fact, the Event Horizon Telescope Collaboration has revealed images of Sagitarius A*'s polarized ring, which provides information about the magnetic fields around the black hole \cite{eht_sgr_A_VII, eht_sgr_A_VIII}. This provides further motivation to study the effects of the electromagnetic interaction in black holes.

Another scenario of interaction between a black hole and electromagnetic fields is the case where a black hole is immersed in an asymptotically uniform magnetic field. It has been shown that external magnetic fields can mimic the magnetic charge of Reissner-Nordström black holes through the study of its ISCO radius \cite{juraeva2021}, meaning that, given an ISCO radius for a black hole with known mass and spin, it is difficult to determine the value of its magnetic charge and the intensity of the magnetic field it is immersed in through this information alone. The effects of an external magnetic field on black holes has been studied by multiple authors \cite{alzahrani2022, aliev2002}
Furthermore, it has been found that when an external magnetic field is applied, a new small region of stable circular orbits appear near the photon sphere of a Schwarzschild black hole  \cite{defelice2003}. This phenomena appears also in alternative theories of gravity \cite{haydarov2020, Haydarov2020cju, rayimbaev2021_EA, rayimbaev2020_EGB}. This work focuses on studying the stable circular orbits of an external asymptotically uniform magnetic field on the properties of a static and spherically symmetric black hole in EdGB gravity.

The present work is organized as follows: Section 2 reviews the static and spherically symmetric black hole solution in EdGB gravity, as well as its photonic orbit. In Section 3, we study the stability of circular orbits in EdGB gravity, around a static and spherically symmetric neutral black hole immersed in an asymptotically uniform magnetic field. In section 4, we compute a few trayectories of particles as example and verification. In Section 5, we present the conclusions of the obtained results.

\section{Theoretical Framework}
\subsection{EdGB Black hole solutions}
In what follows, we use the spacetime signature $(-,+,+,+)$, together with the system of geometrized units, where $G=c=1$. Greek indices take values from $0$ to $3$, while Roman indices go from $1$ to $3$. These conventions will be used in the entirety of this work. Under these considerations, the action in EdGB gravity is given by \cite{kanti1996}:
i
where $R$ is the Ricci scalar, $\Phi$ is the dilaton field, $\alpha$ is a coupling constants, and $R_{GB}^2$ is the Gauss-Bonnet term:
\begin{equation}
    R^2_{GB}=R_{\mu\nu\rho\sigma} R^{\mu\nu\rho\sigma}-4R_{\mu\nu}R^{\mu\nu}+R^2,
\end{equation}
where $R_{\mu\nu\rho\sigma}$ is the Riemann tensor and $R_{\mu\nu}$ is the Ricci tensor. It is more convenient to use a shifted dilaton field $\Psi=\Phi+\ln(\alpha)$ in order to simplify the calculations. Using this action, the resulting field equations for the dilaton and the metric are
\begin{equation}
    \nabla_{\mu} \nabla^{\mu} \Psi = -\frac{1}{4} e^{\Psi} R_{GB}^2,
\end{equation}

\begin{eqnarray}
    G_{\mu\nu}&=&\frac{1}{2}\left(
    \partial_\mu \Psi \partial_\nu \Psi -
    \frac{1}{2} g_{\mu\nu}\partial_\rho \Psi \partial^\rho \Psi
    \right)\nonumber\\
    &&-\frac{1}{4} e^\Psi \left[
    H_{\mu\nu}+4(
    \partial^\rho \Psi \partial^\sigma \Psi + 
    \partial^\rho \partial^\sigma \Psi
    ) P_{\mu\rho\nu\sigma}
    \right],
\end{eqnarray}
where $\nabla_\mu$ is the covariant derivative, $G_{\mu\nu}$ is the usual Einstein tensor, and
\begin{eqnarray}
    H_{\mu\nu}&=&2\left(
    R R_{\mu\nu}-2R_{\mu\sigma} {R^\sigma}_\nu-
    2R_{\mu\rho\nu\sigma}R^{\rho\sigma}+
    R_{\mu\rho\sigma\lambda} {R_\nu}^{\rho\sigma\lambda}
    \right)\nonumber\\
    &&-\frac{1}{2} g_{\mu\nu} R^2_{GB},
\end{eqnarray}

\begin{eqnarray}
    P_{\mu\nu\rho\sigma}=&&R_{\mu\nu\rho\sigma}+
    g_{\mu\sigma}R_{\rho\nu}-g_{\mu\rho}R_{\sigma\nu}+
    g_{\nu\rho}R_{\sigma\mu}-g_{\nu\sigma}R_{\rho\mu}\nonumber\\
    &&+\frac{1}{2}R g_{\mu\rho}g_{\sigma\nu}-\frac{1}{2}R g_{\mu\sigma}g_{\rho\nu}.
\end{eqnarray}

The tensor $H_{\mu\nu}$ also arises in the field equations in Einstein-Gauss-Bonnet gravity. If we evaluate each of its components in four dimensions, we notice that each of its components vanish.  To find a static and spherically symmetric black hole solution, we start from ansatz

\begin{equation}
    \dd s^2 = g_{\mu\nu} \dd x^\mu \dd x^\nu=
    -e^\Gamma \dd t^2 + e^\Lambda \dd r^2 + r^2 (\dd \theta^2 + \sin^2\theta \dd \varphi^2),
\end{equation}
where $\Gamma$ and $\Lambda$ are functions of the radial coordinate.Using a metric of this form and the field equations, it is possible to express $e^{\Lambda}$ in terms of the other unknown functions, then the problem can be reduced to a system of two second-order differential equations:

\begin{equation}
\begin{aligned}
    \Psi''&=\frac{d_1(r,\Lambda, \Gamma', \Psi, \Psi')}{d(r,\Lambda, \Gamma', \Psi, \Psi')}\\
    \Gamma''&=\frac{d_2(r,\Lambda, \Gamma', \Psi, \Psi')}{d(r,\Lambda, \Gamma', \Psi, \Psi')},\\
    e^{\Lambda}&=
    \frac{-Q+\sqrt{Q^2-
    6\Gamma^\prime \Psi^\prime e^\Psi
    }}
    {2},
\end{aligned}
\end{equation}
where 
\begin{equation}
    Q=\frac{{\Psi^\prime}^2r^2}{4}-1-\Gamma^\prime\left(r+
    \frac{e^\Psi \Psi^\prime}{2}\right),
\end{equation}

$d$, $d_1$ and $d_2$ are as defined in the equations (54) to (56) in \cite{kanti1996} (with $\phi$ instead of $\Psi$), and the prime notation is used to indicate differentiation with respect to $r$. These equations can be solved numerically by establishing the usual boundary conditions at the event horizon and, for the dilaton field,
\begin{equation}
    \Psi^\prime_h=r_h e^{-\Psi_h} \left(
    -1+\sqrt{1-6\frac{e^{2\Psi_h}}{r_h^4}}
    \right)
\label{eq:phi_horizon}
\end{equation}
where $\Psi_h$ is the value of the shifted dilaton field at the event horizon, and $r_h$ is the radius of the event horizon. This way, the static and spherically symmetric black hole solutions in EdGB gravity can be parametrized in terms of $\Psi_h$. Furthermore, we notice that existence of $\Psi'_h$ requires that $\Psi_h<\ln(r_h^2/\sqrt{6})$. Also, ensuring that the dilaton field $\Phi$ vanishes at the infinity enables us to recover the value of the coupling constant $\alpha$ with $\alpha=e^{\Psi_\infty}$, where $\Psi_\infty$ is the limit of $\Psi$ as $r$ goes to infinity. We also notice that, from the way the action is defined, the black hole solution approaches the Schwarzschild solution as $\Psi_h \to -\infty$. By considering these restrictions to $\Psi_h$, we find that it is possible to formulate a more convenient parametrization through a parameter $p$ defined as \cite{kokkotas2017}:
\begin{equation}
    p \equiv 6 \frac{e^{2\Psi_h}}{{r_h}^4}=\frac{6{\alpha}^2}
    {{r_h}^4} e^{2\Phi_h}, \quad 0\leq p<1.
\end{equation}

Comparing this parameter with the limits of the value of the dilaton field at the horizon, we notice that the value $p=0$ corresponds to the Schwarzschild solution, while values of $p$ close to 1, respond to stronger couplings. For this work, we use the parameter $p$ to find the black hole solutions numerically with the help of Sagemath. The system of differential equations was solved using an adaptive
step-size fourth-order Runge-Kutta-Nyström algorithm \cite{lund2009track}. 

Furthermore, we compared our results with the approximate analytical solution found in \cite{kokkotas2017} through the use of continued fractions expansion \cite{konoplya2016}. To the fourth order, this solution has proved to be practical, having an error of less than $0.5\%$ for the $(tt)$ compnent of the metric with $p=0.8$, and less than $1.2\%$ for the $(rr)$ component with $p=0.5$. These errors decrease for small values of $p$. When calculating the ISCO radius, the values found with the analytical solution showed an error of less than $1\%$ for values of $p\lesssim 0.95$.

\subsection{Photon orbits in EdGB}
\begin{figure}[hb]
    \centering
    \includegraphics[width=9cm]{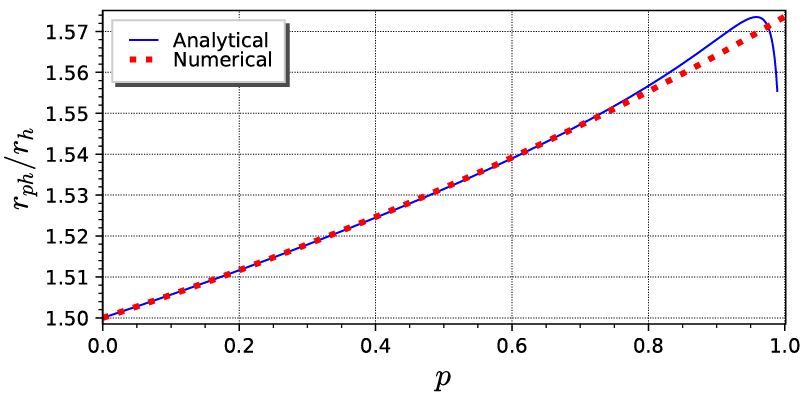}
    \caption{Variation of the radius of photonic orbits with respect to the parameter $p$.}
    \label{fig:rph_p}
\end{figure}

When studying the motion of magnetized particles around a Schwarzschild immersed in an asymptotically uniform magnetic field, it was found that the magnetic field creates a narrow region of possible stable orbits near the photon sphere \cite{defelice2003}. To verify that this phenomenon also holds for EdGB gravity, let us first study the photonic orbits under this theory. Using the definitions $E\equiv p_t$ and $L\equiv p_\varphi$ for the energy and angular momentum of a test particle, we can write the equations of motion this particle as

\begin{equation}
    g_{rr}\left( \dv{r}{\lambda}\right)^2 = -m^2 - \frac{E^2}{g_{tt}}-\frac{L^2}{r^2},
\label{eq:eom1}
\end{equation}
where $\lambda$ is an affine parameter and $m$ is the mass of the particle. For the case of massless particles, we set $m=0$, and rewrite the equation of motion as
\begin{equation}
    \frac{g_{tt} g_{rr}}{L^2}\left( \dv{r}{\lambda} \right)^2=
    -\frac{E^2}{L^2} -\frac{g_{tt}}{r^2}.
\label{eq:pre_photon}
\end{equation}
The conditions for circular orbits are $\dd r /\dd\lambda=0$ and $\dd^2 r/ \dd\lambda^2=0$. By deriving \eqref{eq:pre_photon} with respect to the affine parameter and applying the conditions for circular orbits, we obtain the equation for photonic orbits:
\begin{equation}
    g^\prime_{tt}r-2g_{tt}=0 \quad \Leftrightarrow \quad
    (g_{tt}/r^2)^\prime=0,
\label{eq:photon_equation}
\end{equation}
which can be solved numerically to find the radius of photonic orbits $r_{ph}$. Fig. \ref{fig:rph_p} shows how $r_{ph}$ changes with the parameter $p$ and how the numerical results compare to the ones obtained with the approximate analytical solution. We notice that the value of $r_{ph}$ increases monotonically with respect to $p$. This behavior is similar to that of the radius of innermost stable circular orbits \cite{nampalliwar2018}. The analytical solution deviates minimally up to $p\approx 0.97$ with an error of less than $0.3\%$. Past this point, the curve obtained with the analytical solution proceeds to fall drastically. This plot and \eqref{eq:photon_equation} will let us locate the region of stable orbits that appear when we apply a asymptotically uniform magnetic field to a static and spherically symmetric black hole.

\section{Motion of magnetized particles in a uniform magnetic field}

\subsection{Asymptotically uniform magnetic field}
We study the case of a static and spherically symmetric black hole solution in EdGB gravity inside an asymptotically uniform magnetic field. In \cite{ernst1976}, it was shown that black holes immersed within external magnetic fields are characterized by non-flat solutions for a sufficiently large external magnetic field. However, when the intensity of the magnetic field $B_0$ is sufficiently small ($|B_0 M|\ll\,1$, where $M$ is the black hole mass), the asymptotic black hole solution can be approximately flat and the magnetic field approximately uniform. In this context, we use Wald's procedure \cite{wald1974black}, which assumes the spacetime metric is unperturbed by the magnetic field. Using this method, the electromagnetic four-potential for a black hole with no spin or charge is: 
\begin{equation}
    A_\varphi = \frac{1}{2}B_0 r^2 \sin^2 \theta.
\end{equation}
By using the definition of the anti-symmetric electromagnetic tensor $F_{\mu\nu}=\partial_mu A_\nu - \partial_\nu A_\mu$, the non-vanishing components are 
\begin{equation}
\begin{aligned}
    &F_{r\varphi}=B_0 r \sin^2 \theta\\
    &F_{\theta\varphi}=B_0 r^2 \sin\theta \cos\theta.
\end{aligned}
\end{equation}

We notice that in the equatorial plane, the only non vanishing component of the electromagnetic tensor is $F_{r\varphi}$.

\subsection{Circular orbits}
The Hamilton-Jacobi equation for a magnetized particle can be written as \cite{defelice2003}:
\begin{equation}
    g^{\mu\nu} \pdv{S}{x^\mu} \pdv{S}{x^\nu}=
    -\left(
    m-\frac{1}{2} D^{\mu\nu} F_{\mu\nu}
    \right)^2,
\label{eq:hamilton-jacobi}
\end{equation}
where $S$ is the action of the magnetized particle in the curved spacetime and $D_{\mu\nu}$ is the polarization tensor, which contains the electromagnetic properties of the particle. The product $D^{\mu\nu}F_{\mu\nu}$ represents the electromagnetic interaction between the particle and the external magnetic field. For a particle with no electric charge, the polarization tensor can be expressed as \cite{defelice2003}
\begin{equation}
    D^{\mu\nu} \equiv \eta^{\mu\nu\rho\lambda} u_\rho \mu_\lambda,
    \quad D^{\mu\nu} u_\nu=0,
\label{eq:polarization}
\end{equation}
where $u^{\mu}$ is the four-velocity of the particle, and $\mu^\lambda$ is the four-vector of its magnetic dipole moment. The electromagnetic tensor allows the 3+1 decomposition
\begin{equation}
    F_{\alpha \beta}=u_{[\alpha} E_{\beta]}-
    \eta_{\alpha\beta\sigma\gamma} u^\sigma B^\gamma.
\end{equation}
This decomposition, together with \eqref{eq:polarization}, leads to
\begin{equation}
    D^{\mu\nu}F_{\mu\nu}=
    2\mu^{\hat{\alpha}} B_{\hat{\alpha}}=
    2\mu B_0 \mathcal{L}[\lambda_{\hat{\alpha}}],
\label{eq:magnetic_product}
\end{equation}
where $\mu=|\vec{\mu}|=\left(\mu^{\hat{\alpha}}\mu_{\hat{\alpha}}\right)^{1/2}$ is the module of the dipole magnetic moment of the magnetized particle, the hatted indexes represent the projection of the components onto an orthonormal tetrad frame $\{\lambda_{\hat{\alpha}}\}$ adapted to the fiducial observer, and $\mathcal{L}[\lambda_{\hat{\alpha}}]$ is a function of the spacetime coordinates and the parameters that define the tetrad $\{\lambda_{\hat{\alpha}}\}$. Furthermore, $\mu^{\hat{\alpha}}$ reduces to the classical magnetic dipole moment vector $\vec{\mu}$ measured by a comoving observer in the tetrad frame. We will assume the magnetic dipole moment of the particle is aligned with the external magnetic field.

For simplicity, we consider the case of weak magnetic interaction, so we can approximate $\left(D^{\mu\nu}F_{\mu\nu}\right)\to 0$. The action of the magnetized particle can be expressed as
\begin{equation}
    S=-Et+L\varphi + S_r.
\end{equation}
Using this action in \eqref{eq:hamilton-jacobi}, we obtain the equation of radial motion of the magnetized particle:

\begin{equation}
    -g_{tt}\ g_{rr} \left(\dv{r}{\tau}  \right)^2+V_{\text{eff}}=\varepsilon^2,
\label{eq:eom_particle}
\end{equation}
where we defined the effective potential as
\begin{equation}
    V_{\text{eff}}=-g_{tt}-
    l^2\frac{g_{tt}}{r^2}+\beta\ g_{tt}\ \mathcal{L}[\lambda_{\hat{\alpha}}],
\label{eq:Veff}
\end{equation}
$l=L/m$ is the specific angular momentum, and $\beta=2\mu B_0/m$ is called the magnetic coupling parameter. This definition of effective potential allows us to write the conditions for circular orbits as 
\begin{equation}
    \dv{r}{\tau}=0, \quad \pdv{V_{\text{eff}}}{r}=0.
\label{eq:conditions_veff}
\end{equation}

From the first condition for $\beta$, we obtain
\begin{equation}
    \beta(r;l,\varepsilon, p)=\frac{1}{\mathcal{L}[\lambda_{\hat{\alpha}}]}
    \left(1+\frac{\varepsilon^2}{g_{tt}}+
    \frac{l^2}{r^2}
    \right),
\label{eq:eq_beta_pre}
\end{equation}

Using both conditions for circular orbits, and comparing to the derivative of $\beta(r;l,\varepsilon, p)$,
\begin{equation}
    \pdv{V_{\text{eff}}}{r}=\mathcal{L}[\lambda_{\hat{\alpha}}] \pdv{r} \beta
    (r;l,\varepsilon, p).
\label{eq:dVeff_dbeta}
\end{equation}

Now, it is necessary to find an expression for $\mathcal{L}[\lambda_{\hat{\alpha}}]$ to be able to solve the equations of motion of the magnetized particle. To do this, we assume that the particle is moving along the equatorial plane. For this circular motion, the tetrad of a fiducial comoving observer can be expressed as \cite{defelice1992,defelice2003}
\begin{equation}
\begin{aligned}
\lambda_{\hat{0}}=&e^{\psi} \partial_t+
\Omega e^{\psi}\partial_{\varphi}\\
\lambda_{\hat{r}}=&\left[
-\Omega  \sqrt{-\frac{g_{\varphi\varphi}}{g_{tt}}}\partial_t-
\sqrt{-\frac{g_{tt}}{g_{\varphi\varphi}}}\partial_{\varphi} \right]e^{\psi} \sin \left(\Omega_{FW}t\right)\\
&+\sqrt{\frac{1}{g_{rr}}}\cos \left(\Omega_{FW}t\right) \partial_r\\
\lambda_{\hat{r}}=&\sqrt{\frac{1}{g_{\theta\theta}}}\partial_\theta\\
\lambda_{\hat{\varphi}}=&\left[
\Omega  \sqrt{-\frac{g_{\varphi\varphi}}{g_{tt}}}\partial_t+
\sqrt{-\frac{g_{tt}}{g_{\varphi\varphi}}}\partial_{\varphi} \right]e^{\psi} \cos \left(\Omega_{FW}t\right)\\
&+\sqrt{\frac{1}{g_{rr}}}\sin \left(\Omega_{FW}t\right) \partial_r,
\end{aligned}
\end{equation}
where $e^\psi=(-g_{tt}-r^2\Omega^2)^{-1/2}$, $\Omega=\displaystyle\frac{\dd\varphi}{\dd r}=-\frac{g_{tt}}{r^2}\frac{l}{\varepsilon}$ is the angular velocity of the orbiting particle, and $\Omega_{FW}$ is the Fermi-Walker angular velocity. We can now find the components of the electromagnetic tensor in the new tetrad:
\begin{equation}
    B_{\hat{r}}=B_{\hat{\varphi}}=0,
    \quad B_{\hat{\theta}}=B_0 e^\psi
    \sqrt{\frac{-g_{tt}}{g_{rr}}}.
\label{eq:magnetic_tetrad}
\end{equation}
Let us remember that the magnetic dipole moment of the particle is aligned with the external magnetic field for the case we are studying. Taking this into account, \eqref{eq:magnetic_product} and \eqref{eq:magnetic_tetrad} give
\begin{equation}
    \mathcal{L} [\lambda_{\hat{\alpha}}]=
    \sqrt{\frac{g_{tt}}{g_{rr}\left(g_{tt}+r^2 \Omega^2 \right)}}=\left[ g_{rr}\left(1+\frac{l^2}{\varepsilon^2} \frac{g_{tt}}{r^2}\right)\right]^{-1/2}
\end{equation}

which, together with \eqref{eq:eq_beta_pre}, gives
\begin{equation}
    \beta(r;l,\varepsilon, p)=
    \left[ g_{rr}\left(1+\frac{l^2}{\varepsilon^2} \frac{g_{tt}}{r^2}\right)\right]^{1/2}
    \left(1+\frac{\varepsilon^2}{g_{tt}}+
    \frac{l^2}{r^2}
    \right)
\label{eq:beta_function}
\end{equation}

\begin{figure}[htb!]
    \centering
    \begin{subfigure}[b]{0.48\textwidth}
        \centering
        \includegraphics[width=\textwidth]{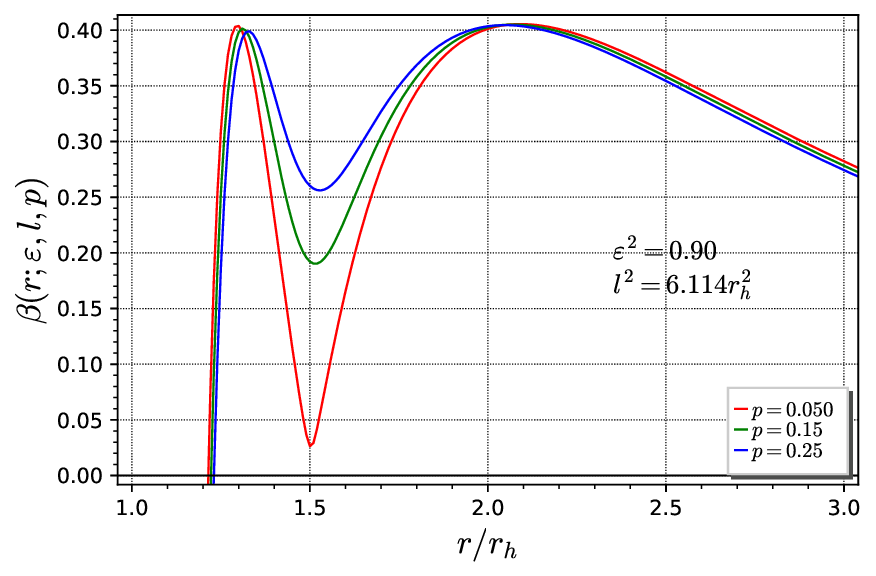}
        \caption{}
        \label{fig:fbet_a}
    \end{subfigure}
    \hfill
    \begin{subfigure}[b]{0.48\textwidth}
        \centering
        \includegraphics[width=\textwidth]{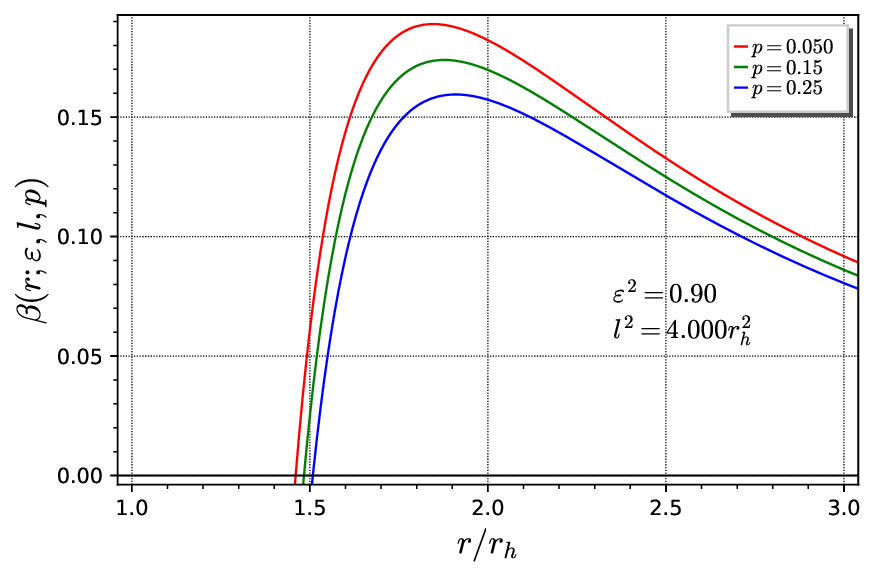}
        \caption{}
        \label{fig:fbet_b}
    \end{subfigure}
\caption{Plots of the $\beta(r;l,\varepsilon,p)$ for different values of $p$ and $l$.}\label{fig:fbet}
\end{figure}

Fig. \ref{fig:fbet} shows the dependence of $\beta(r;l,\varepsilon,p)$ with respect to $r$ for $\varepsilon^2 = 0.90$ and a pair of values of $l^2/r_h^2$. The plots show that $\beta(r;l,\varepsilon,p)$ presents a minimum near the photonic orbit, similar to other gravity theories. This minimum is more pronounced for lower values of $p$ and higher values of $l$. For low values of $l$, it is shown that the increase of the parameter $p$ reduces the maximum value of the magnetic coupling parameter $\beta$ that allows circular orbits. On the other hand, for high values of $l$, the maxima of $\beta(r;l,\varepsilon, p)$ has little change for different values of $p$, while the local minimum near the photonic sphere increases as $p$ increases. Furthermore, a reduction of the specific angular momentum seems to have a similar effect on $\beta(r;l,\varepsilon, p)$ as an increase in the parameter $p$.

Having an expression for $\mathcal{L}[\lambda_{\hat{\alpha}}]$, it is easy to see that this function should always be non-negative. Therefore, \eqref{eq:dVeff_dbeta} implies that $\partial V_{\text{eff}}/\partial r$ and $\partial \beta (r;l,\varepsilon, p)/\partial r$ will have the same sign. This allows us to rewrite the conditions for circular orbits as:
\begin{equation}
    \beta= \beta(r;l,\varepsilon, p),\quad
     \pdv{r}\beta(r;l,\varepsilon, p)=0
\label{eq:orbits_system_2}
\end{equation}

Here we have a system of two equations with five unknwown parameters ($\beta, r, l, \varepsilon, p$), so its solution
can be expressed in terms of any three of five independent
variables. It is more convenient to use the magnetic coupling parameter $\beta$, the radius of the circular orbits $r$, and the dimensionless parameter $p$ as free parameters.

\subsection{Stability of circular orbits}

The second equation in \eqref{eq:orbits_system_2} can be arranged as a quadratic equation with respect to the square of the specific energy $\varepsilon^2$, which has the solutions

\begin{equation}
    \varepsilon^2_{extr}=\frac{
    g'_{rr}g_{tt}r^8-g_{rr}^3l^2\left(r^6 g_{tt}/g_{rr}^2\right)'+\sigma r^8\sqrt{\Delta_\varepsilon}}
{-2r^8g_{tt}^2\left(g_{rr}/g_{tt}^2\right)'}
\label{eq:e_extr}
\end{equation}
where $\sigma = \pm 1$ and

\begin{equation}
\begin{aligned}
    \Delta_\varepsilon=&
    \left(g'_{rr}g_{tt}\right)^2 +2l^2g_{rr}\left(g_{tt}/r^2\right)'\left(4g'_{tt}g_{rr}+g'_{rr}g_{tt}\right)+\\&
    \left[3l^2 g_{rr}(g_{tt}/r^2)'\right]^2.
\label{eq:delta_eps}
\end{aligned}
\end{equation}

The existence of $\varepsilon_{extr}$ depends on the positivity of $\Delta_\varepsilon$. For the case $p=0$, which reduces to the Schwarzschild solution, $\Delta_\varepsilon$ is a perfect square, making it positive for all values of $l$ and $r$. To study what happens for other values of $p$, we rewrite $\Delta_\varepsilon$ as
\begin{equation}
    \Delta_\varepsilon=\left[3g_{rr}\left(\frac{g_{tt}}{r^2}\right)^\prime\right]^2(l^2-l^2_{lim+})(l^2-l^2_{lim-}),
\label{eq:delta_eps2}
\end{equation}
with

\begin{equation}
\begin{aligned}
    l^2_{lim\pm}=\frac{ 
    ( g_{tt}^4 g_{rr})^\prime\pm
    2\sqrt{(g_{tt} g_{rr})^3 (g_{tt}^4/g_{rr}^2)^\prime (g_{tt} g_{rr})^{\prime}}}{-9 g_{tt}^3\, g_{rr} \left(g_{tt}/r^2\right)^\prime}.
\end{aligned}
\label{eq:l2lim}
\end{equation}

\begin{figure*}[t!]
     \centering
     \begin{subfigure}[b]{0.49\textwidth}
         \centering
         \includegraphics[width=\textwidth]{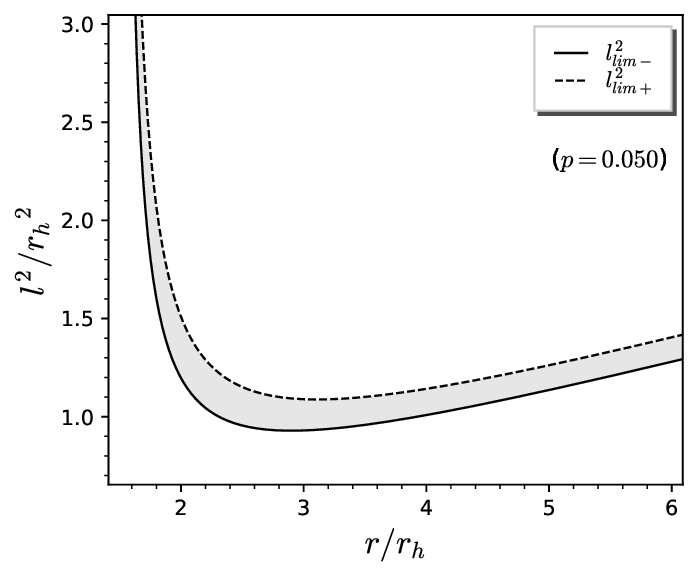}
         \caption{}
         \label{fig:l2lim_p05}
     \end{subfigure}
     \hfill
     \begin{subfigure}[b]{0.49\textwidth}
         \centering
         \includegraphics[width=\textwidth]{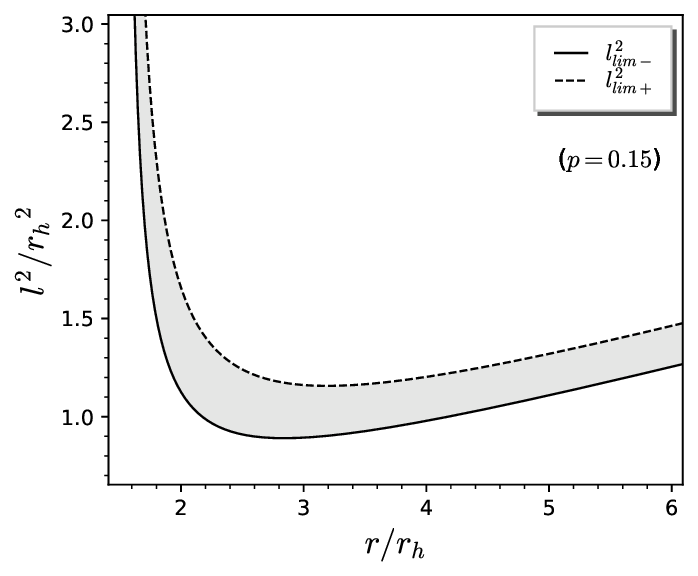}
         \caption{}
         \label{fig:l2lim_p15}
     \end{subfigure}
     \hfill
     \begin{subfigure}[b]{0.49\textwidth}
         \centering
         \includegraphics[width=\textwidth]{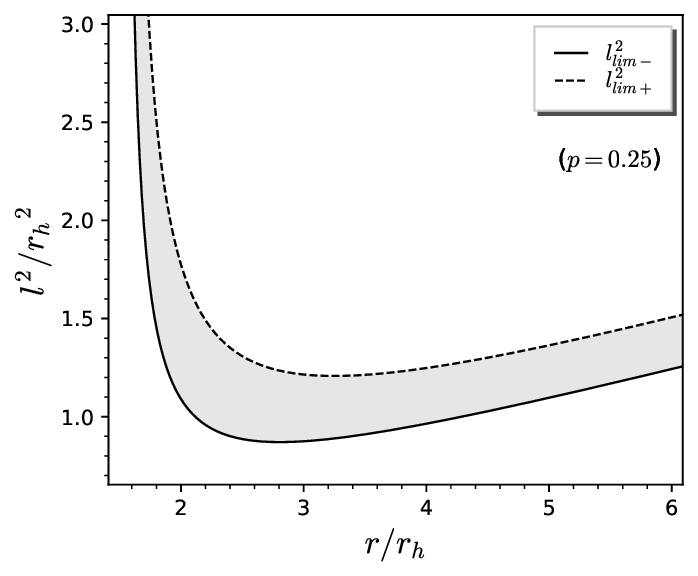}
         \caption{}
         \label{fig:l2lim_p25}
     \end{subfigure}
     \hfill
     \begin{subfigure}[b]{0.49\textwidth}
         \centering
         \includegraphics[width=\textwidth]{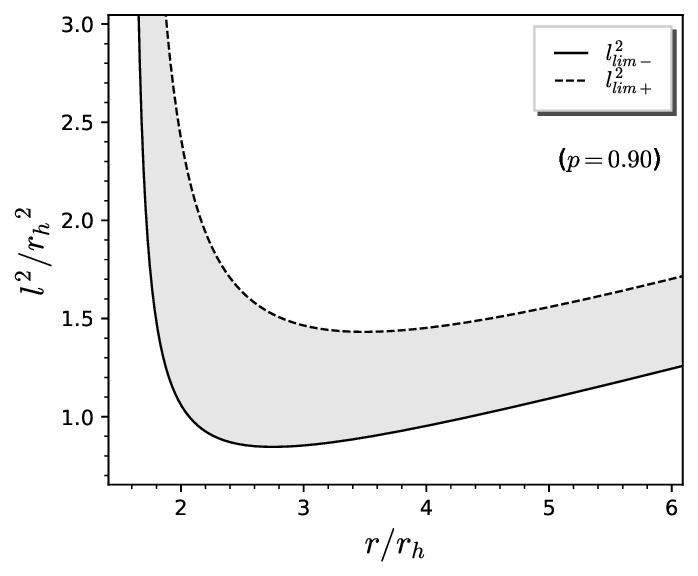}
         \caption{}
         \label{fig:l2lim_p90}
     \end{subfigure}
\caption{Plot of the functions $l^2_{lim\pm}$ for different values of $p$. The gray area represents the forbidden regions where no circular orbits are possible. \subref{fig:l2lim_p05} $p=0.05$. \subref{fig:l2lim_p15} $p=0.15$. \subref{fig:l2lim_p25} $p=0.25$. \subref{fig:l2lim_p90} $p=0.90$.}
\label{fig:l2lim}
\end{figure*}

We notice that the sign of $\Delta_\varepsilon$ depends on the last two factors in \eqref{eq:delta_eps2}, therefore, for a fixed value of $r$, the value of the angular momentum $l$ defines the positivity of $\Delta_\varepsilon$, which is required for the existence of $\varepsilon_{extr}$. In order to ensure the positivity of $\Delta_{\varepsilon}$, it is required that the value of $l^2$ does not lie between $l^2_{lim-}$ and $l^2_{lim+}$. It is possible to show from \eqref{eq:l2lim} that $l^2_{lim-}<l^2_{lim+}$ outside the photon sphere, which implies that $\Delta_{\varepsilon}$ is positive when either $l^2<l^2_{lim-}$ or $l^2>l^2_{lim+}$. This means that, for a given value of $l^2$, such that $l^2_{lim-}<l^2<l^2_{lim+}$, no circular orbits are possible for any given values of the  coupling parameters $(\beta,p)$.

To visualize the dependence of $l^2_{lim\pm}$ with respect to the coordinate $r$, we plot these functions on Fig. \ref{fig:l2lim} for different values of $p$. In these graphics, we see that, for a fixed value of $l^2$, there could be up to two regions where no circular orbits are allowed, while some small values of $l^2$ can avoid these forbidden regions. We also notice that, as we reduce the value of $p$, the forbidden region becomes smaller until it becomes nonexistent for $p=0$. In general, \eqref{eq:l2lim} implies that this forbidden region disappears ($l^2_{lim+}=l^2_{lim-}$) for solutions where $g_{tt} \propto 1/g_{rr}$. We also notice that \eqref{eq:photon_equation} implies that $l_{lim\pm}$ increases asymptotically as we approach the photonic sphere. Henceforth, we are going to focus on the regions where $l^2 \leq l^2_{lim-}$, since we are interested in the innermost circular orbits.

\begin{figure*}[htb!]
     \centering
     \begin{subfigure}[b]{0.49\textwidth}
         \centering
         \includegraphics[width=\textwidth]{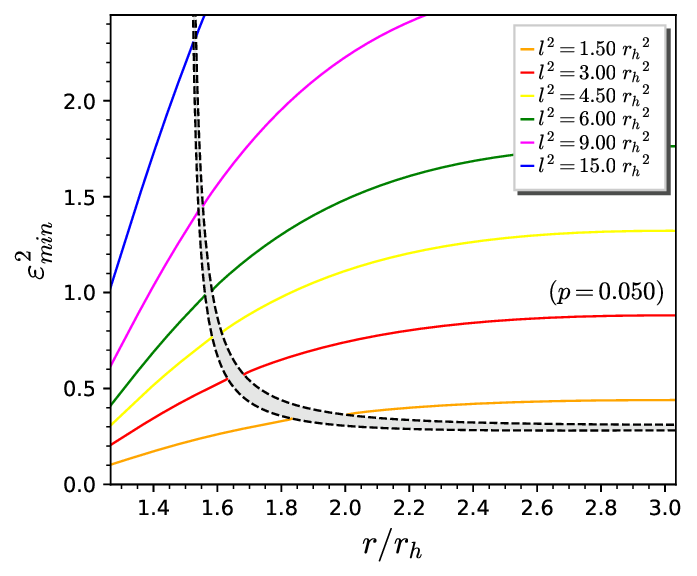}
         \caption{}
         \label{fig:e2min_p05}
     \end{subfigure}
     \hfill
     \begin{subfigure}[b]{0.49\textwidth}
         \centering
         \includegraphics[width=\textwidth]{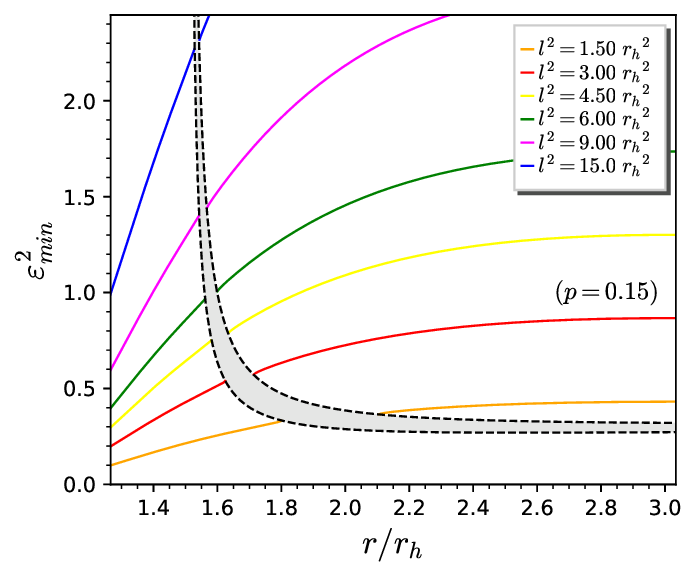}
         \caption{}
         \label{fig:e2min_p15}
     \end{subfigure}
     \hfill
     \begin{subfigure}[b]{0.49\textwidth}
         \centering
         \includegraphics[width=\textwidth]{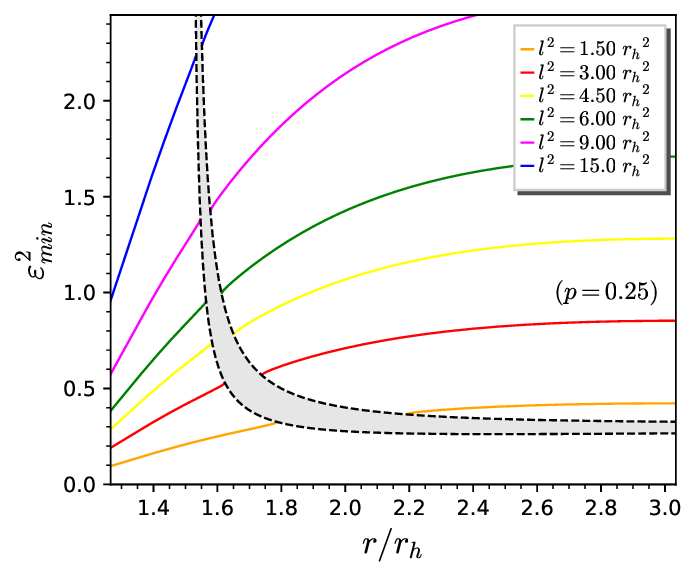}
         \caption{}
         \label{fig:e2min_p25}
     \end{subfigure}
     \hfill
     \begin{subfigure}[b]{0.49\textwidth}
         \centering
         \includegraphics[width=\textwidth]{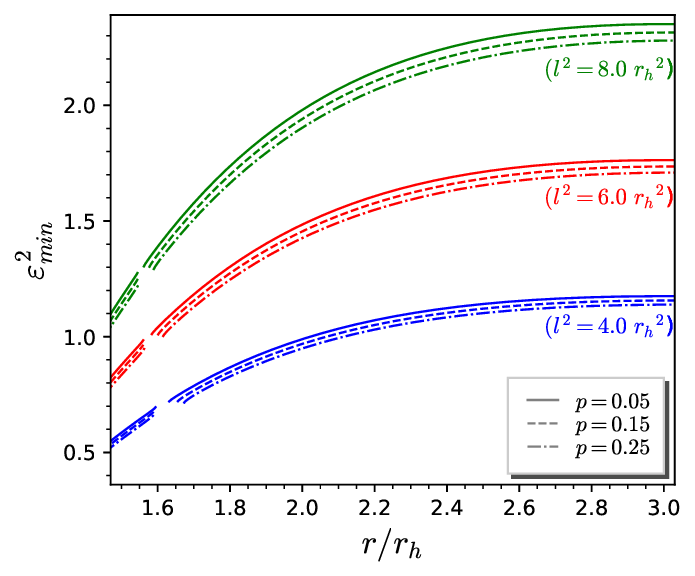}
         \caption{}
         \label{fig:e2min_lp}
     \end{subfigure}
\caption{Plots of $\varepsilon^2_{min}$ for different values of $l^2$ and $p$. The gray area represents the forbidden regions where no circular orbits are possible, which appear as discontinuities on \subref{fig:e2min_lp}.}
\label{fig:e2min}
\end{figure*}

Let us now study how this forbidden region affects the possible values of energy $\varepsilon^2_{extr}$ corresponding to a extremum of $\beta(r;l\varepsilon,p)$. The curves that delimit the forbidden region for the energy can be obtained by inserting \eqref{eq:l2lim} into \eqref{eq:e_extr}. The value of $\sigma$ becomes irrelevant to find the limiting values of $\varepsilon^2_{extr}$ because both limiting values of $l^2$ makes $\Delta_\varepsilon$ vanish in \eqref{eq:e_extr}. On the other hand, for a fixed value of $l^2$, we need to determine which value of $\sigma$ corresponds to a minimum of $\beta(r;l\varepsilon,p)$. Through a careful computer analysis, we found that, near the photonic sphere, the branch with $\sigma=-1$ corresponds to the minimum of $\beta(r;l\varepsilon,p)$, meanwhile, after crossing the forbidden region, the branch switches to $\sigma=+1$ to keep generating a minimum. That is, for $r>r_{ph}$:
\begin{equation}
    \varepsilon^2_{min}=
    \begin{cases} 
       \varepsilon^2_{extr}\big\rvert_{\sigma=-1}, & l^2_{lim-}>l^2
       \\
      \varepsilon^2_{extr}\big\rvert_{\sigma=+1}, & l^2_{lim+}<l^2 \\
   \end{cases}
\label{eq:e_min}
\end{equation}

Fig. \ref{fig:e2min} shows the dependence of $\varepsilon^2_{min}$ with respect to $r$ for different values of $l^2$, and how it behaves before and after going thtough the forbidden region. We can see that $\varepsilon^2_{min}$ increases with $l^2$, and that the forbidden region becomes wider as we increase the value of the EdGB parameter $p$. We also notice that $\varepsilon^2_{min}$ decreases with the increase of the parameter $p$, as shown in Fig. \ref{fig:e2min}\subref{fig:e2min_lp}, and the discontinuities in each curve show the regions where no real value for $\varepsilon^2_{min}$ can be found.

The value of $\beta(r;l,\varepsilon,p)$ at its minimum can be obtained by inserting the corresponding expression for $\varepsilon^2$ into \eqref{eq:beta_function}:
\begin{equation}
    \beta_{min}(r;l,p)=\beta(r;l,\varepsilon_{min},p).
\label{eq:betamin}
\end{equation}

\begin{figure*}[htb!]
     \centering
     \begin{subfigure}[b]{0.49\textwidth}
         \centering
         \includegraphics[width=\textwidth]{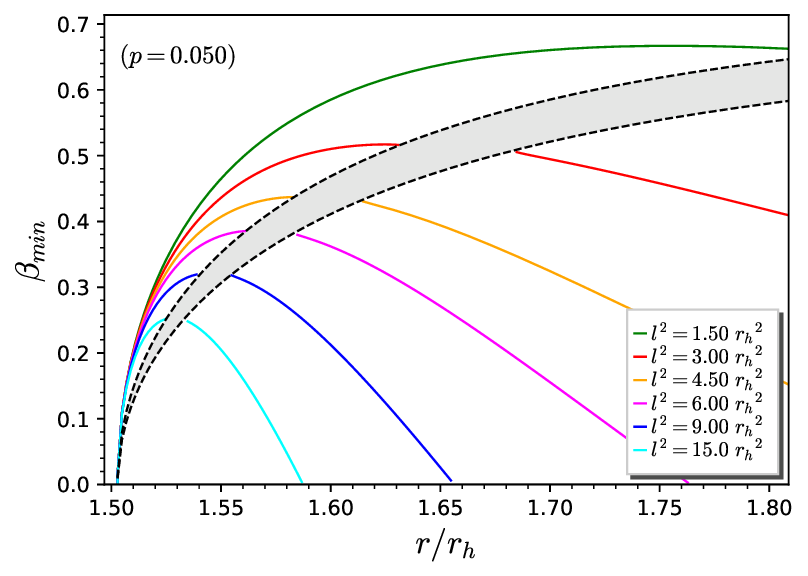}
         \caption{}
         \label{fig:bmin_p05}
     \end{subfigure}
     \hfill
     \begin{subfigure}[b]{0.49\textwidth}
         \centering
         \includegraphics[width=\textwidth]{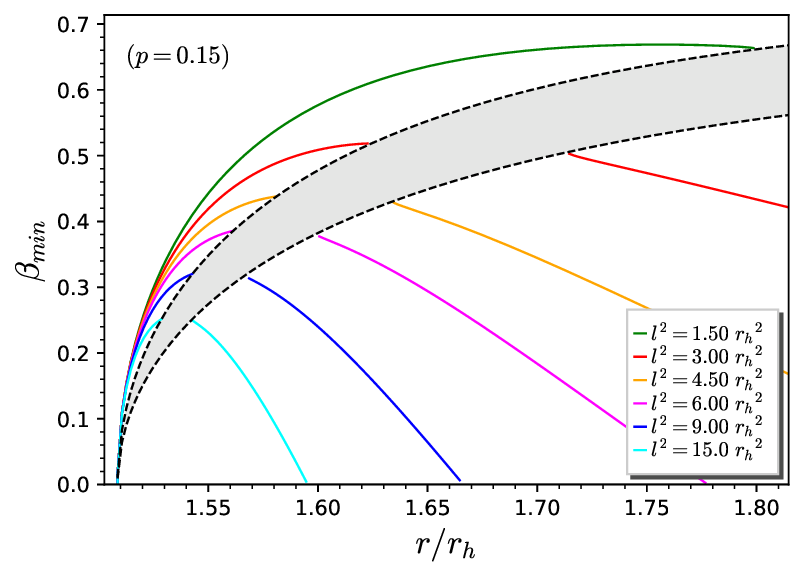}
         \caption{}
         \label{fig:bmin_p15}
     \end{subfigure}
     \hfill
     \begin{subfigure}[b]{0.49\textwidth}
         \centering
         \includegraphics[width=\textwidth]{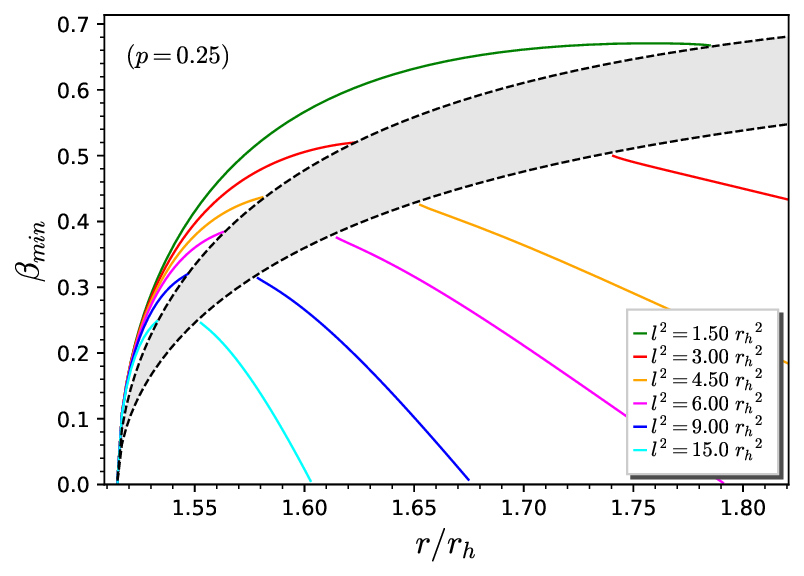}
         \caption{}
         \label{fig:bmin_p25}
     \end{subfigure}
     \hfill
     \begin{subfigure}[b]{0.49\textwidth}
         \centering
         \includegraphics[width=\textwidth]{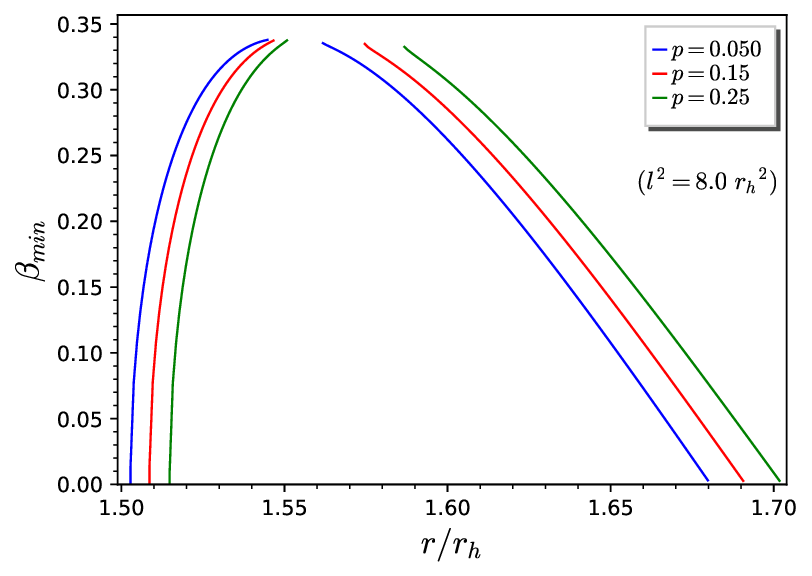}
         \caption{}
         \label{fig:bmin_l_r}
     \end{subfigure}
\caption{Plot of the functions $\beta_{min}$ for different values of $p$. The gray area represents the forbidden regions where no circular orbits are possible, which appear as discontinuities on \subref{fig:bmin_l_r}.}
\label{fig:bmin}
\end{figure*}

Similar to how it happens with $l^2$ and $\varepsilon^2_{min}$, $\beta_{min}(r;l,p)$ also encounters forbidden regions. The limits of these forbidden regions are found by inserting \eqref{eq:l2lim} into \eqref{eq:betamin}:
\begin{equation}
    \beta_{lim\pm} (r;p)= \beta_{min}(r;l_{lim\mp},p).
\label{eq:blim}
\end{equation}

This means that, given a value of $r$, there are no possible circular orbits for $\beta_{lim-}<\beta<\beta_{lim+}$. We now study the value of $\beta_{min}$ near the photonic orbit. Using the photonic orbit equation \eqref{eq:photon_equation} in \eqref{eq:delta_eps}, \eqref{eq:e_min} and \eqref{eq:beta_function}, we find that, for any value of angular momentum, $\beta_{min}$ vanishes at $r_{ph}$
\begin{equation}
    \left. \beta_{min} \right\rvert_{r=r_{ph}}=0,
\end{equation}

which marks the limit for stable circular orbits. Fig. \ref{fig:bmin} shows the dependence of $\beta_{min}(r;l,p)$ with respect to the coordinate $r$ for different values of $l^2$. We observe how, consistently with the cases for $l^2$ and $\varepsilon^2_{min}$, the forbidden region for $\beta_{min}(r;l,p)$ gets wider with the increase of the EdGB parameter $p$. We also see how for different values of $p$, an increase of the angular momentum $l^2$ results in a decrease of $\beta_{min}(r;l,p)$, but in general, all the curves converge to $0$ at $r_{ph}$ as it was found by using the photonic orbit equation \eqref{eq:photon_equation}. We can also see that an increase of the parameter $p$ causes a ``shift" of the curve of $\beta_{min}$ to the right.

In the case of the Schwarzschild solution, to find the maximum value for stable circular orbits, we can solve $\partial\beta_{min} / \partial r = 0$ for $l^2$, which would give us the minimum value of the specific angular momentum $l^2_{min}$, corresponding to the minimum value of the magnetic coupling parameter. Inserting $l^2_{min}$ into the expression for $\beta_{min}$ will give us the value of the magnetic coupling parameter $\beta_{extr}$ for which the circular orbit with radius $r$ is marginally stable \cite{defelice2003}. Nevertheless, there are a pair of difficulties that arise when trying to apply this process to EdGB black holes. The first one being the complexity of the analytical expression of $\beta_{min}$ and its derivative, and the second one arises due to the existence of the forbidden regions that we have found.

Graphically, $l_{min}$ tells us the value of the angular momentum that causes $\beta_{min}$ to have a maximum at a distance $r$ from the origin. By looking at Fig. \ref{fig:bmin}, we notice that, for large values of $l^2$, there's no visible maximum of $\beta_{min}$, implying that $\partial \beta_{min}/\partial r=0$ has no solution for large values of $l^2$ and values of $r$ close to the photonic sphere. Therefore, we are interested in finding the mimum value of $r$ that allows a maximum of $\beta_min$, which we call $r_{crit}$. First notice from Fig. \ref{fig:bmin} that, at the photonic sphere, $\beta_{min}$ is increasing; if we find the conditions for $\beta_{min}$ to be decreasing when entering the forbidden region, these will be the conditions for the existence of local maximum. Let us assume a fixed value of $l^2$; as $r$ increases, $\beta_{min}$ will approach the forbidden region until it reaches it at $r=r_F$. If we evaluate the derivative of $\varepsilon^2_{min}$ right before entering the forbidden region ($\sigma= -1,\, \Delta_{\varepsilon}=0$), Eqs. \eqref{eq:e_min} and \eqref{eq:e_extr} imply
\begin{equation}
\begin{aligned}
\pdv{r}\varepsilon^2_{min}=
\frac{1}{4 g_{tt}^2(g_{rr}/g_{tt}^2)^\prime} \frac{\Delta_{\varepsilon}^\prime}{\sqrt{\Delta_{\varepsilon}}}+
\begin{pmatrix}
\text{terms finite}\\
\text{at } r_F\end{pmatrix}.
\end{aligned}
\end{equation}

Since $r_F$ marks the entrance to the forbidden region, this is where $\Delta_{\varepsilon}$ goes from being positive to negative, meaning that, at $r_F$, we have $\Delta_{\varepsilon}=0$ and $\Delta_{\varepsilon}^\prime>0$. Furthermore, due to the black hole boundary conditions, $g_{rr}/g_{tt}^2$ is a decreasing function that goes from $+\infty$ at the event horizon and goes to $1$ at the infinity. Taking all of this into consideration, we find the lateral limit
\begin{equation}
    \lim_{r\to {r_F}^-}{\pdv{r}\varepsilon^2_{min}}=+\infty.
\end{equation}

Since, $\partial\beta_{min}/\partial r$ depends on $\partial \varepsilon^2_{min}/\partial r$, it will also diverge at the boundary of the forbidden region. It is possible to show that this is the only divergent term in $\partial\beta_{min}/\partial r$, thus, the derivative of $\beta_{min}$ can be expressed as
\begin{equation}
\begin{aligned}
    \pdv{r}\beta_{min}=&
    \frac{g_{rr}\mathcal{L}[\lambda_{\hat{\alpha}}]}{2\varepsilon_{min}^2} 
    \Sigma(r;p,l)\pdv{r}\varepsilon^2_{min}+\dots
\end{aligned}
\end{equation}

where 
\begin{equation}
\begin{aligned}
    \Sigma(r;p,l)=
    \left(1+
    \frac{\varepsilon_{min}^2}{g_{tt}}+
    \frac{l^2}{r^2} \right) 
     \left(2-\frac{l^2}{\varepsilon_{min}^2}\frac{g_{tt}}{r^2}\right)-
    2
\end{aligned}
\end{equation}
and the rest of the terms are finite at $r_F$. Due to the divergent behavior of $\partial \varepsilon_{min}^2/\partial r $ as we approach the forbidden region, the sign of $\partial \beta_{min}/\partial r$ will be determined by the sign of $\Sigma (r;p,l)$ at $r_F$, where $l^2=l^2_{lim-}$, 

\begin{figure}[htb!]
    \centering
    \includegraphics[width=9cm]{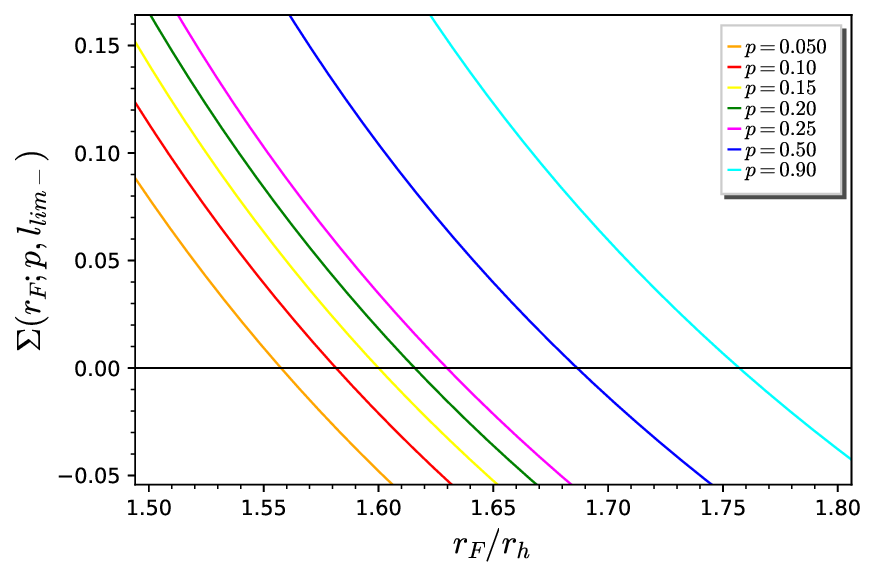}
    \caption{Plot of $\Sigma(r_F;p, l_{lim-})$ for different values of $p$. The regions where $\Sigma(r_F;p, l_{lim-})>0$ indicate the regions where $\beta_{min}$ will not have a local maximum before entering the forbidden region.}
    \label{fig:sign_beta}
\end{figure}

\begin{figure}[htb]
    \centering
    \includegraphics[width=9cm]{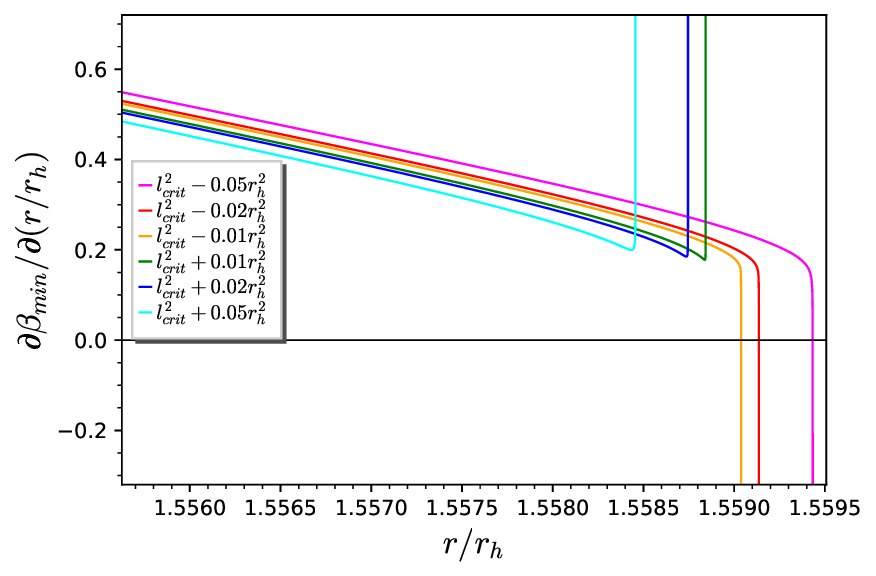}
    \caption{Plot of $\partial\beta_{min}/\partial(r/r_h)$for values of $l^2$ close to $l^2_{crit}$, and $p=0.05$, using the approximate analytical solution.}
    \label{fig:l2crit}
\end{figure}

Fig. \ref{fig:sign_beta} shows the behavior of $\Sigma (r_F;p, l_{lim-})$; the negativity of this function indicates the regions where maxima of $\beta_{min}$ are possible. We identify the values of $r_F$ where  $\Sigma (r_F;p, l_{lim-})=0$ as the minimum orbit radius $r_{crit}$ that allows a maximum for $\beta_{min}$. We can see that $r_{crit}$ increases as $p$ increases, which is consistent with the increase of width of the forbidden region portrayed in Fig. \ref{fig:bmin}. If one performs this analysis with the approximate analytical solution,  it can be verified tha the values of $r_{crit}$ tend to be slightly larger than the ones found with the numerical solution. Furthermore, the maximum angular momentum $l^2_{crit}$ that allows the existence of the maximum of the magnetic coupling parameter $\beta_{extr}$ can be found by evaluating $l^2_{lim-}$ at $r_F$ in \eqref{eq:l2lim}. To illustrate this, Fig. \ref{fig:l2crit} shows an example of how $\partial \beta_{min}/\partial r$ behaves for values of angular momentum close to $l^2_{crit}$. We notice how values of $l^2$ smaller than $l^2_{crit}$ allow roots of $\partial \beta_{min}/\partial r$ as they decrase asymptotically. Conversely, values of angular momentum greater than $l^2_{crit}$ cause an asymptotic increase, making it impossible for maxima of $\beta_{min}$ to exist. The minimum value of $\beta_{extr}$ can be found by inserting $r_{crit}$ into $\beta_{lim-}$, since this point corresponds to the intersection of both functions.
\begin{equation}
    \beta_{crit}(p)=\beta_{lim-}(r_{crit}; p)
\end{equation}
We can conclude that, for $\beta<\beta_{crit}$, the maximum radius of stable circular orbits $r_{max}$ can be found by solving $\beta=\beta_{lim-}$ for $r$. When $\beta>\beta_{crit}$, $r_{max}$ can be found by solving $\partial\beta_{min}/\partial r=0$ and $\beta=\beta_{min}$ numerically.

To find the minimum radius of stable circular orbits $r_{min}$ for a given value of $\beta$, we evaluate $\beta_{min}$ for $l=0$. It can be verified that $\varepsilon^2_{min}$ becomes $0$ for $l=0$, which can lead to an indeterminate form in \eqref{eq:beta_function}. For this reason, we evaluate the limit of $l^2/\varepsilon^2_{min}$ as $l^2\to 0$:

\begin{equation}
    \lim_{l^2 \to 0} {\frac{l^2}{\varepsilon^2_{min}}}=
    \frac{-g^{\prime}_{rr}}{\left(g_{tt}g_{rr}/r^2\right)^\prime}.
\end{equation}

\begin{figure*}[htb!]
    \centering
     \begin{subfigure}[b]{0.47\textwidth}
         \centering
         \includegraphics[width=\textwidth]{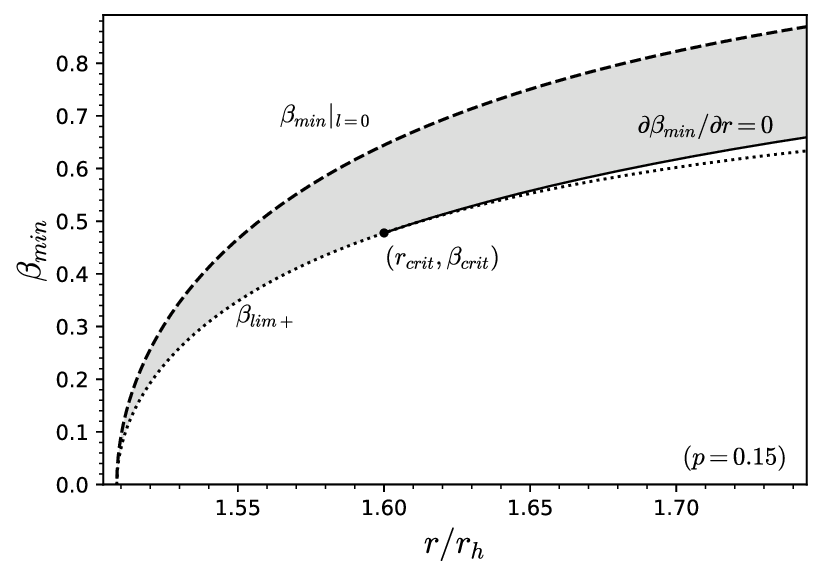}
         \caption{}
         \label{fig:stable_diagram}
     \end{subfigure}
     \hfill
     \begin{subfigure}[b]{0.49\textwidth}
        \centering
        \includegraphics[width=\textwidth]{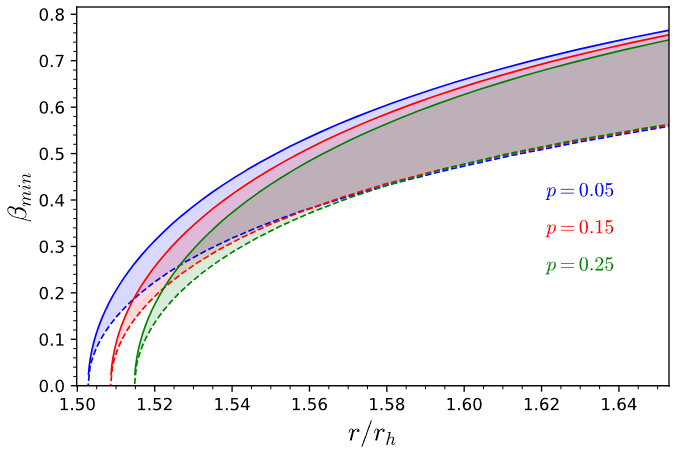}
        \caption{}
        \label{fig:stable_orbits}
     \end{subfigure}
\caption{ \subref{fig:stable_diagram} Diagram showing the criteria to determine the region of stable orbits. The dotted line marks the frontier of the forbidden region. \subref{fig:stable_orbits} Stable orbits for different values of the parameter $p$.}
\label{fig:stable}
\end{figure*}

Inserting this result into \eqref{eq:betamin}, with $l=0$, we get

\begin{equation}
    \beta_{min}(r;p)\big\rvert _{l=0} = g_{rr}\sqrt{\frac{\left(g_{tt}/r^2\right)^\prime}{\left(g_{tt}g_{rr}/r^2\right)^\prime}}
\label{eq:bminl0}
\end{equation}
Then, one can find the minimum radius of circular stable orbits by solving $\beta_{min}(r_{min};l,p)\rvert _{l=0}=\beta$ for $r_{min}$, which can be performed numerically. Using these criteria, we found the region of allowed stable circular orbits numerically, which corresponds to values of $r$ between $r_{min}$ and $r_{max}$. \ref{fig:stable_diagram} shows a diagram that explains visually how the upper and lower bounds of stable orbits are determined using $p=0.15$ as example. As it was explained, the curves that mark the upper limit for stable circular orbits are determined by $b_{lim+}$ when $r<r_{crit}$ and by the local maxima of $\beta_{min}$ for $r>_{crit}$, nevertheless, the change between the two regions of the curve are too small to perceive visually. \ref{fig:stable_orbits} shows the regions of stable orbits for different values of $p$, where we can notice how these regions are pushed outwards as $p$ increases, and how its width decreases with the increase of $p$. We are interested in having an expression for the width of the region of allowed stable circular orbits $\Delta r = r_{max}-r_{min}$ but the complexity of the equations makes it difficult to obtain analitically. Instead, we show the values of $\Delta r$ for different values of $p$ on Table \ref{tab:delta_r}. In the table, we notice how the value of $\Delta r$ changes subtly with the increase of $p$, but it is consistently decreasing, reducing the region of stable circular orbits.

\begin{table*}[htb!]
\centering
\begin{tabular}{p{2cm} p{2cm}  p{2cm}  p{2cm}  p{2cm}  p{2cm}}
    \hline
    $p $  & $\beta=0.05$ & $\beta=0.10$ & $\beta=0.20$ & $\beta=0.30$ & $\beta=0.50$ \\
    \hline
    $0.00$ & $0.00052$ & $0.0021$ & $0.0086$ & $0.0204$ & $0.0674$ \\
    $0.05$ & $0.00040$ & $0.0016$ & $0.0068$ & $0.0168$ & $0.0632$ \\
    $0.10$ & $0.00035$ & $0.0014$ & $0.0060$ & $0.0150$ & $0.0591$  \\
    $0.15$ & $0.00032$ & $0.0013$ & $0.0055$ & $0.0137$ & $0.0551$ \\
    $0.20$ & $0.00030$ & $0.0012$ & $0.0051$ & $0.0127$ & $0.0513$ \\
    $0.25$ & $0.00028$ & $0.0011$ & $0.0048$ & $0.0119$ & $0.0480$ \\
    $0.30$ & $0.00026$ & $0.0011$ & $0.0045$ & $0.0116$ & $0.0452$ \\
    $0.40$ & $0.00023$ & $0.0009$ & $0.0040$ & $0.0100$ & $0.0407$ \\
    $0.50$ & $0.00021$ & $0.0009$ & $0.0037$ & $0.0091$ & $0.0372$ \\
    $0.70$ & $0.00018$ & $0.0007$ & $0.0031$ & $0.0078$ & $0.0322$ \\
    \hline
\end{tabular}
\caption{ Values of $\Delta r/r_h$ for different values of $p$ and $\beta$, calculated numerically.}
\label{tab:delta_r}
\end{table*}

\begin{figure}[hb]
    \centering
    \includegraphics[width=9cm]{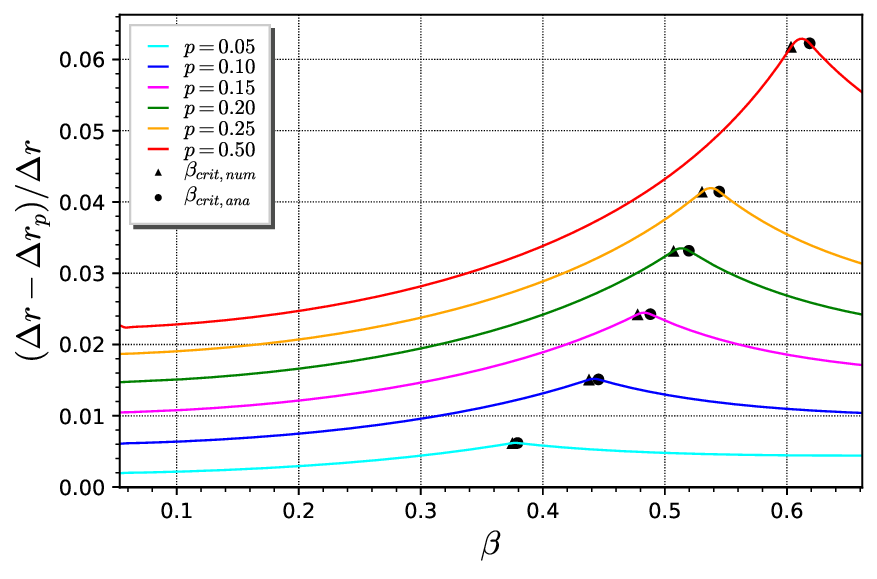}
    \caption{Error of $\Delta r$ found with the approximate analytical solution with respect to the numerical solutions.}
    \label{fig:Delta_r}
\end{figure}

Using this procedure, we used the approximate analytical solution to find how $\Delta r$ deviates from the numerical calculations. In Fig. \ref{fig:Delta_r}, we show the percent error found for $\Delta r$ for multiple values of $p$. It is shown that the error reaches a maximum where the values of $\beta_{crit}$ found numerically and analytically differ, which is to be expected. Furthermore, the maximum relative error stays below $5\%$ for $p<0.25$.

Using the graphic on Fig. \ref{fig:stable_orbits}, for a given value of $p$, one can check if a pair $(\beta, r)$ lies within the region of allowed circular stable orbits. The angular momentum for that case can be found by solving $\beta = \beta_{min}(r;l,p)$ numerically for $l$, and then insert it on \eqref{eq:e_min} to find the energy of the particle for a specific circular orbit. An alternative approach is using the substitution $\zeta=l/\varepsilon$ on the equation of motion \eqref{eq:eom_particle}. The conditions for circular orbits $\dd r/\dd \tau=0$ and $\partial V_{\text{eff}}/\partial r=0$ become
\begin{equation}
    \frac{
    r^2\left(
    1-\beta \mathcal{L}[\lambda_{\hat{\alpha}}]
    \right)}{r^2+\zeta^2\ g_{tt}}
    \left(
    \frac{g_{tt}^\prime}{g_{tt}}+\frac{2\zeta^2 g_{tt}}{r^3}
    \right)-\beta \pdv{r} \mathcal{L}[\lambda_{\hat{\alpha}}]=0,
\label{eq:eq_zeta}
\end{equation}
\begin{equation}
    \varepsilon^2=\frac{- r^2 g_{tt}}{r^2+\zeta^2 g_{tt}}\left(
    1-\beta \mathcal{L}[\lambda_{\hat{\alpha}}]
    \right), \quad l^2=\varepsilon^2 \zeta^2.
\label{eq:e2l2_circ}
\end{equation}

Given $p$, $\beta$ and $r$, \eqref{eq:eq_zeta} can be solved numerically for $\zeta^2$, which then can be used to find $\varepsilon^2$ and $l^2$ from \eqref{eq:e2l2_circ}.

One might be curious about how the forbidden region behaves far from the photon sphere. Even though we have studied how these regions obstruct the stable orbits near the photon sphere, these can affect outer orbits as well. For this, we use the expansions \cite{kanti1996}:
\begin{equation}
\begin{aligned}
g_{tt}&=-1+\frac{2M}{r}+\mathcal{O}(1/r^3)\\
g_{rr}&=1+\frac{2M}{r}+
\frac{4M^2-D^2}{r^2}+\mathcal{O}(1/r^3),
\end{aligned}
\end{equation}
on the expression for $l_{lim\pm}$ \eqref{eq:l2lim} and then into \eqref{eq:e_extr}, which shows that, for large values of $r$, the limits of the forbidden region for the specific angular momentum and energy behave as 
\begin{equation}
\begin{aligned}
l^2_{lim\pm}&=\frac{Mr}{3}+\mathcal{O}(\sqrt{r})\\
\varepsilon^2_{lim\pm}&=\frac{1}{3} + \mathcal{O}(1/\sqrt {r}),
\end{aligned}
\end{equation}
meaning that both, $\varepsilon^2_{lim+}$ and $\varepsilon^2_{lim-}$ converge to the value of $1/3$ at infinity, whereas $l^2_{lim\pm}$ approach to a straight line as $r$ goes to infinity. This implies the forbidden region gets smaller as the particle gets far from the black hole. This result is valid for any value of the EdGB parameter $p$.

\subsection{Some particle trayectories}

Having obtained the equations of motion of a test particle, it is possible to find the trayectory that this particle would follow around a static and spherically symmetric black hole immersed in an asymptotically uniform magnetic field by studying the shape of the effective potential. Nevertheless, the effective potential defined in \eqref{eq:Veff} depends of the energy of the particle, so any change on the energy of the particle would change the shape of the effective potential, making it difficult to find the limits for bound orbits. For the purpose of the study of particle trayectories, and compare different cases, we use $\zeta=l/\varepsilon$, to rewrite the equation of motion \eqref{eq:eom_particle} as

\begin{equation}
    {-g_{rr}g_{tt}}\left(1+\zeta^2 \frac{g_{tt}}{r^2}\right)^{-1}\left(
    \dv{r}{\tau}{}\right)^2+\mathcal{V}_{\zeta}=\varepsilon^2,
\end{equation}
where
\begin{equation}
    \mathcal{V}_\zeta=-g_{tt}\left(1+\zeta^2 \frac{g_{tt}}{r^2}\right)^{-1}+\beta \frac{g_{tt}}{\sqrt{g_{rr}}}\left(1+\zeta^2 \frac{g_{tt}}{r^2}\right)^{-3/2}.
\end{equation}

\begin{figure}[ht!]
    \centering
    \includegraphics[width=0.5\textwidth]{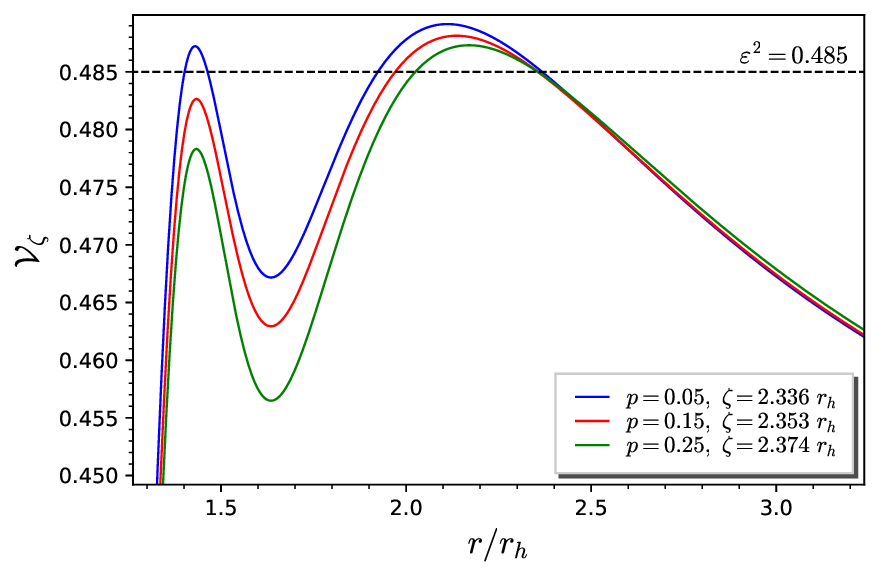}
    \caption{Effective potential $\mathcal{V}_{\zeta}$ with $\beta=0.55$ for different values of $p$. The values of $\zeta$ were chosen for each case such that they allow circular orbits around $r/r_h=1.635$.}
    \label{fig:Vzeta}
\end{figure}

\begin{figure*}[ht!]
     \centering
     \begin{subfigure}[b]{0.32\textwidth}
        \centering
        \includegraphics[width=\textwidth]{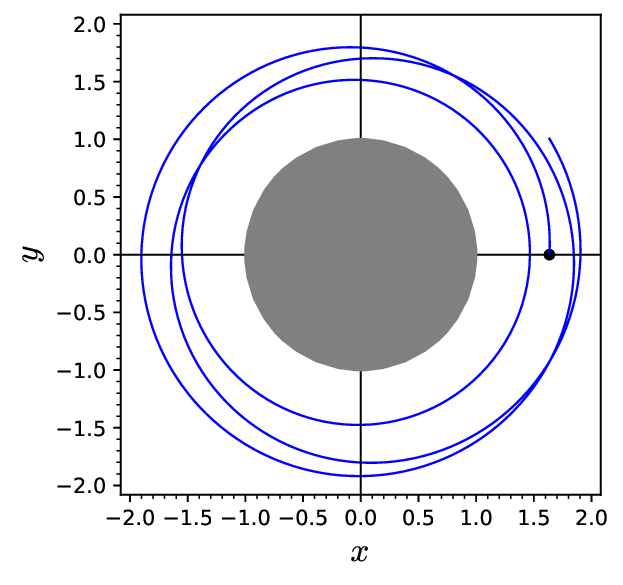}
        \caption{$p=0.05$}
        \label{fig:trayp05}
     \end{subfigure}
     \hfill
     \begin{subfigure}[b]{0.32\textwidth}
        \centering
         \includegraphics[width=\textwidth]{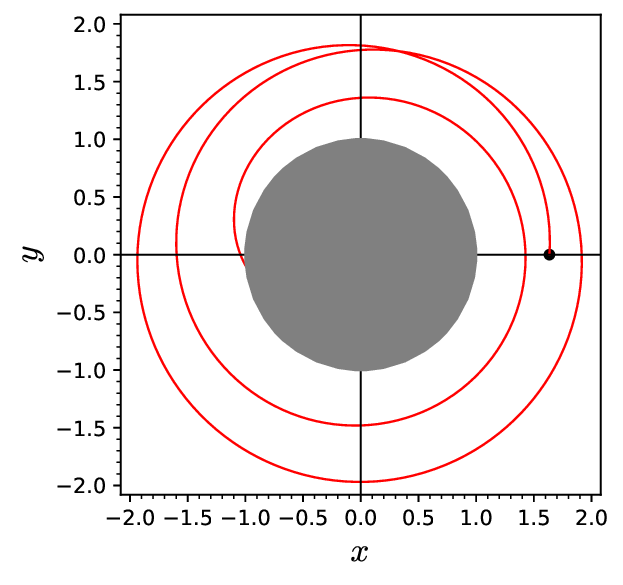}
         \caption{$p=0.15$}
         \label{fig:trayp15}
     \end{subfigure}
     \hfill
     \begin{subfigure}[b]{0.32\textwidth}
        \centering
         \includegraphics[width=\textwidth]{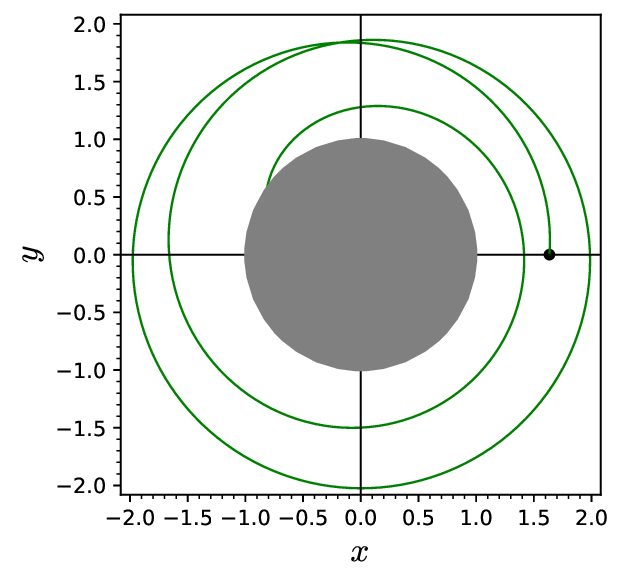}
         \caption{$p=0.25$}
         \label{fig:trayp25}
     \end{subfigure}
     \caption{Trayectories of a magnetized particle with specific energy $\varepsilon^2=0.485$ around a static and spehrically symmetric black hole in EdGB gravity. }
     \label{fig:tray}
\end{figure*}

This way, we can compare the energy of the particle with the new effective potential as long as we assume a fixed ratio $l/\varepsilon$. As example, we consider the cases where $p=0.05, 0.15$ and $0.25$ as illustrated on Fig.\ref{fig:stable_orbits}, choosing a region where all three cases allow stable circular orbits. We choose a magnetic coupling parameter $\beta=0.55$ for orbits around $r/r_h=1.635$. Using these parameters, we solved \eqref{eq:eq_zeta} numerically and plotted the effective potentials $\mathcal{V}_\zeta$ shown in Fig. \ref{fig:Vzeta}. From this plot, we notice that a value of specific energy $\varepsilon^2=0.485$ should allow bound orbits for $p=0.05$ only. For all three cases, increasing the energy (while keeping the ratio $l/\varepsilon$ constant) will eventually cause the particle to fall towards the black hole. This fall happens first for higher values of $p$. The shape of the effective potential also tells us that the region between the periastron and apoastron tends to be bigger as we increase the value of $p$. The trayectories for these cases were obtained by solving
\begin{equation}
\begin{aligned}
        \dv{t}{\tau}&=\frac{\varepsilon}{-g_{tt}}\\
        \left(\dv{r}{\tau}\right)^2&=\frac{1}{g_{rr}}\left(-1-\frac{\varepsilon^2}{g_{tt}}-
        \frac{l^2}{r^2}+
        \beta
        \mathcal{L}[\lambda_{\hat{\alpha}}]\right)\\
        \dv{\varphi}{\tau}&=\frac{l}{r^2}.
\end{aligned}
\end{equation}

The numerical solutions shown in Fig, \ref{fig:tray} verify the behavior of the test particle as predicted from the effective potential. For the given values of $\zeta$ and $\beta$, the particle describes bound orbits for $p=0.05$, and falls into the black hole for $p=0.15$ and $p=0.25$. This can lead us to think that orbits in this region are more stable for lower values of $p$, as it was reflected in the width of the regions of stable orbits shown in Fig. \ref{fig:stable_orbits}.

\section{Conclusion}
We studied the motion of magnetized particles around a static and spherically symmetric black hole embedded in an asymptotically uniform magnetic field for EdGB gravity. Similar to other alternative theories, the magnetic interaction creates a region near the photonic sphere that allows the existence of circular stable orbits. The radius of the photonic sphere increases with the value of the EdGB parameter $p$, which means that the region of stable circular orbits is pushed away with the increase of $p$. It was found that for small values of specific angular momentum $l$, the maximum value of the coupling magnetic parameter $\beta$ that allows circular orbits decreases as the value of $p$ increases. For high values of $l$, the local minimum of $\beta$ that allows circular orbits near the photonic sphere increases as $p$ increases.

The unsual form of the metric of the static and spherically symmetric black hole solution in EdGB leads to some regions where no stable circular orbits are possible. These regions restrict the possible values of angular momentum $l$, energy $\varepsilon$ and magnetic coupling parameter $\beta$. We also notice that this forbidden region gets wider as the value of $p$ increases. For large values of $r$, the forbidden region disappears. For circular orbits, an increase of the angular momentum increase the value of the energy of the orbiting particle $\varepsilon$ and decreases the value of the magnetic coupling parameter $\beta$. An increase of the EdGB parameter $p$ decreases the value of $\varepsilon$ slightly and "pushes" the plot of $\beta$ vs $r$ to the right due to the increase of the radius of photonic sphere.

When studying the stability of circular orbits, we found that there's a value $\beta_{crit}$ such that, when the magnetic coupling parameter $\beta$ is smaller than $\beta_{crit}$, the stable orbits are limited by the forbidden region. By restricting the forbidden regions in the calculations procedure, we found the regions of stable circular orbits near the photonic sphere for different values of $p$, numerically. For a fixed value of $\beta$, width $\Delta r$ fo these regions decrease as the value of $p$ increases. These regions are also pushed away from the event horizon with the increase of $p$, due to how the photonic sphere changes. When comparing the width of the regions of stable circular orbits with the approximate analytical solution to the fourth order, we found that the error reaches a maximum near $\beta_{crit}$, and these maxima stay below $5\%$ for $p<0.25$.

Finally, the trayectories of magnetized particles around a static and spherically symmetric EdGB black hole immersed in an asymptotically uniform magnetic field show that these orbits tend to be more unstable for higher values of $p$. Also, an increase of energy of the orbiting test particle, while keeping the ration $l/\varepsilon$ constant, will tend to make it fall towards the black hole before getting enough energy to escape from it.

\end{document}